\documentstyle[array,multicol,aps,preprint]{revtex}
\tightenlines

\newcommand{\eref}[1]{(\ref{#1})}

\begin{document}
\preprint{{hep-th/0204139} \hfill {UCVFC-DF-25-2002}}
\title{Link Invariants from Classical Chern-Simons Theory}
\author{Lorenzo Leal{${}^{a, b, }$}\footnote{lleal@fisica.ciens.ucv.ve}}
\address{${}^a$Grupo de Campos y Part\'{\i}culas, Departamento de F\'{\i}sica, Facultad de Ciencias, Universidad
Central de Venezuela, AP 47270, Caracas 1041-A, Venezuela \footnote{permanent address}\\
${}^b$Departamento de F\'{\i}sica Te\'orica, Universidad
Aut\'onoma de Madrid, Cantoblanco 28049, Madrid, Spain}
%

\maketitle

\begin{abstract}
Taking as starting point a perturbative study of the classical
equations of motion of the non-Abelian Chern-Simons theory with
non dynamical sources, we obtain  analytical expressions for link
invariants. In order to present these expressions in a manifestly
diffeomorphism-invariant form, we introduce a set of differential
forms associated with submanifolds in $R^3$, that are metric
independent, and that allow us to consider the link invariants as
a kind of surface-dependent diffeomorphism invariants that present
certain Abelian gauge symmetry.

\end{abstract}

\section{Introduction}

Since the discovery of the relation between Quantum Field Theories
and Knot Theory, there has been an important progress both from
the physical and the mathematical points of view. The starting
point of this interplay was the recognition of the vacuum
expectation value of the Wilson Loop in the Chern-Simons theory
as a polynomial invariant of knots\cite{witten}. Soon, the
perturbative study of the Wilson Loop average, using standard
Feynmann rules, showed that every term in the perturbative series
produces a knot invariant too\cite{guadagnini}.

 As it was pointed out in a recent review\cite{labastida}, interesting issues coming from
the physical side have their mathematical counterpart in
Chern-Simons theory. For instance,  gauge freedom is related with
the fact that there exist different representations for knot
invariants, corresponding to different gauge fixings. All of
these developments have been performed within the Quantum Field
Theory framework.

Recently, the author sketched a proposal to study link invariants
from classical non-Abelian Chern-Simons theory\cite{lorenzo1},
which is based on a previous work about the Abelian
case\cite{lorenzo2}. That preliminary proposal was incomplete in
at least to aspects. First, the action taken as starting point
was not gauge invariant. Then, the results were forced to be
gauge invariant by imposing a consistence condition $ad \,\,
hoc$. Secondly, there was no $a\,\,posteriori$ checking of the
diffeomorphism invariance of the results obtained. The purpose of
this article is to elaborate further on that program, and to
remedy those aspects. To this end, we consider the classical
equations of motion for the non-Abelian Chern-Simons field
coupled to particles carrying non abelian charge (Wong
particles)\cite{wong} , and argue that the on-shell action of
this model should lead to analytical expressions for link
invariants of the world lines of the particles. The action that
we take is due to Balachandran $et \,\,al$ \cite{balachandran},
and, unlike the action that we employed in reference
\cite{lorenzo2}, it is gauge invariant. In view of the non
linearity of the system, which prevents us to obtain exact
solutions of the equations of motion, we develop a perturbative
scheme to solve them. From it we explicitly obtain the first two
contributions to the on-shell action. It is found that they
correspond to the Second and Third Milnor Linking Coeficients,
which are the first two in a family of link invariants of
increasing complexity discovered by Milnor (the Second Coeficient
coincides with the Gauss Linking Number)\cite{milnor}. We also
provide diffeomorphism covariant expressions for the link
invariants obtained. To this end, we introduce a set of
differential forms associated with volumes, surfaces, paths and
points in $R^3$. These forms are easy to manipulate and also
allow for a simple geometric interpretation of the link
invariants obtained.

The paper is organized as follows. In section \ref{sec2} we
present the model. In section \ref{sec3} we discuss the method for
obtaining  link invariants from the solution of the classical
equations of motion. Also, we consider the consistence conditions
that the perturbative equations must obey in order to preserve
gauge invariance. In section \ref{sec4} we introduce the
differential forms mentioned above, and show that the link
invariants obtained by our method can be cast into a manifestly
 diffeomorphism invariant form. Some final remarks are left
for the last section.

\section{Chern-Simons-Wong Theory}\label{sec2}

Our starting point will be the  action

\begin{equation}
S=S_{CS}+S_{int},\label{accion}
\end{equation}
where

\begin{equation}
S_{CS}=-\Lambda^{-1}\int  d^3x \,  \epsilon^{\mu\nu\rho} \,
Tr(A_\mu
\partial_\nu A_\rho + \frac 23 A_\mu A_\nu A_\rho)\label{accion c s}
\end{equation}
is the $SU(N)$ Chern-Simons action and

\begin{equation}
S_{int}= \sum_{i=1}^n \int_{\gamma_{i}}d\tau \, Tr(K_{i}
g^{-1}_{i}(\tau)D_{\tau}g_{i}(\tau))\label{accion int}
\end{equation}
corresponds to the  interaction, through the Chern-Simons field,
between $n$ particles of Wong, which are classical particles
carrying non-abelian charge\cite{wong,balachandran}. We take
$Tr(T^{a}T^{b})=- \frac 12 \delta^{ab}$ and $T^{a}T^{b}=
f^{abc}T^{c}$ for the $N^{2}-1$ generators $T^{a}$ of the $SU(N)$
algebra. We shall use the notation $A_\mu = A_\mu^{a}T^{a} $ ,
$A_{i}(\tau)=A_\mu(z_{i}(\tau))\dot{z}^{\mu}_{i}(\tau)$. In eq.
\eref{accion int}, the curve $\gamma_{i}$ represents the world
line of the $i-th$ particle.  The $SU(N)$ matrix $g_i (\tau)$ is a
dynamical variable, from which one can construct the
chromo-electric charge $I_{i}(\tau)$ as

\begin{eqnarray}
I_{i}(\tau)&\equiv & g_{i}(\tau) K_{i}g^{-1}_{i}(\tau)\nonumber\\
& = &I_{i}^{a}(\tau)T^{a},\label{def I}
\end{eqnarray}
where $K_{i}\equiv K_{i}^{a}T^{a}$ is a constant element of the
algebra, which, as we shall see, is related to the initial value
of the chromo-electric charge $I_{i}(\tau)$. In eq. \eref{accion
int} also appears the covariant derivative of $g_{i}(\tau)$ along
the world line of the $i-th$ particle

\begin{equation}
D_{\tau}g_{i}(\tau)= \dot{g}_{i}(\tau) + A_{i}(\tau)g_{i}(\tau).
\label{deriv covar}
\end{equation}

The dynamical variables are the gauge potentials  $A^{a}_\mu$ and
the matrices $g_{i}(\tau)$ associated with the internal degrees of
freedom of the Wong particles. One could also add to the action
the usual  contribution of the free particles

\begin{equation}
S_{particles}= -\int{d \tau \sum_{i}\sqrt{\mid
\dot{z}^{\mu}_{i}(\tau)\mid^2}} \label{accion part},
\end{equation}
and  consider the trajectories $z^{\mu}_{i}(\tau)$ as dynamical
objects too. However, this is not convenient for our purposes,
since we want to take the curves $\gamma_{i}$ as external objects
whose linking properties  are going to be studied. More
precisely, we shall seek for link invariants related to closed
curves in $R^3$. We shall take these curves just as the world
lines of the Wong particles, which will follow externally
prescribed trajectories. Furthermore, observe that in absence of
the term given by eq.\eref{accion part}, the action that we take
is topological, and this property is necessary in our program. It
should be noticed that the relation of these particles with
physically propagating ones is only formal, because, since we are
interested in knotting properties of curves, we shall take the
particles world lines as closed curves in Euclidean three space.

The Chern-Simons action is invariant under gauge transformations
connected with the identity

\begin{equation}
 A_\mu \to A^{\Omega}_\mu = \Omega^{-1} A_\mu \Omega + \Omega^{-1}\partial_ \mu  \Omega  . \label{gauge t}
\end{equation}
On the other hand, the action $S_{int}$ is also gauge invariant
provided that

\begin{equation}
 K_{i}^{\Omega}=K_{i},\label{gauge k}
\end{equation}

\begin{equation}
 g_{i}^{\Omega}=\Omega^{-1}g_{i},\label{gauge g}
\end{equation}
which in turn implies

\begin{equation}
 (D_{\tau}g_{i})^{\Omega}=\Omega^{-1}D_{\tau}g_{i} ,\label{gauge Dg}
\end{equation}
as corresponds to a covariant derivative. In view of the above
equations, the non-Abelian charge transforms gauge-covariantly in
the adjoint representation

\begin{equation}
 I_{i}^{\Omega}=\Omega^{-1}I_{i} \Omega .\label{gauge I}
\end{equation}

Varying the action \eref{accion} with respect to $A_\mu$ we
obtain the field equation

\begin{equation}
\epsilon^{\mu\nu\rho}F_{\nu\rho}= \Lambda J^{\mu},\label{CS
ecuacion}
\end{equation}
where the current is given by

\begin{equation}
J^{\mu}(x) = \sum_{i=1}^{n} \int_{\gamma_i}d\tau\,
\dot{z}^{\mu}_{i}(\tau) \, I_{i}(\tau)\,
\delta^{3}(x-z_{i}(\tau)),\label{corriente}
\end{equation}
and the field strength is $F_{\mu\nu} = \partial_{\mu}A_{\nu} -
\partial_{\nu}A_{\mu} + \,[A_{\mu},A_{\nu}]$.
On the other hand, to vary the action with respect to  the
$SU(N)$ elements $g(\tau)$, one must isolate the independent
degrees of freedom. This can be accomplished by parametrizing the
group elements as\cite{balachandran}

\begin{equation}
g((\xi(\tau)) = \exp (\xi^{a}(\tau)T^{a}), \label{parametrizacion}
\end{equation}
and then varying the action with respect to the $N^{2}-1$ (for
each particle) independent variables $\xi^{a}$ . The resulting
Euler-Lagrange equations

\begin{equation}
\frac{\partial L}{\partial \xi^{a}_{i}(\tau)} -
\frac{d}{d\tau}\Big(\frac{\partial L}{\partial
\dot\xi^{a}_{i}(\tau)}\Big) =0  , \label{euler lagrange}
\end{equation}
where

\begin{equation}
L \equiv \sum_{i} Tr (K_i g^{-1}_{i}(\tau) D_{\tau}g_{i}(\tau)) ,
\label{parametrizacion}
\end{equation}
can be seen to be equivalent to gauge-covariant conservation of
the non-Abelian charge of each particle along its world
line\cite{balachandran}
\begin{equation}
D_{\tau}I_{i}= \dot{I}_{i} + [A_{i},I_{i}]=0. \label{conservacion}
\end{equation}
The solution of this equation can be written as
\begin{equation}
I_{i}({\tau})= U_{i}({\tau}) \, I_{i}(0)\,U^{-1}_{i}({\tau}) ,
\label{solucion I}
\end{equation}
where $U_{i}({\tau})$ is the time ordered exponential of the
gauge potential along the world line $\gamma_{i}$

\begin{equation}
U_{i}({\tau})= T exp \,\,( -\int_{0}^{\tau} A_{i}({\tau}')\,
d\tau' \,\,) . \label{tiempo ord}
\end{equation}
It can be seen that  equation \eref{conservacion}, on the other
hand, is necessary to fulfill  the consistence condition that
rises by taking the covariant derivative on both sides of equation
\eref{CS ecuacion}. Summarizing, we have that the
Chern-Simons-Wong model described by the action  \eref{accion} is
self-consistent, gauge-invariant and, if the world lines of the
particles are externally given, also topological, in the sense
that it is metric-independent.

\section{Link Invariants}\label{sec3}

Let us suppose that we are able to solve the non linear equations
of motion \eref{CS ecuacion}  under suitable boundary conditions,
obtaining the gauge fields as functionals of the curves
$\gamma_i$. Then, the on-shell action will be expressed  in terms
of these curves. But, since the action $S$ is a metric-independent
scalar function, the same will hold for the on-shell action
$S_{os}$. Therefore, $S_{os}$ has to be a metric-independent
functional of curves, i.e., a link invariant. This should be
compared with what occurs in the quantum evaluation of the vacuum
expectation value of the Wilson Loop
$W(C)$\cite{witten,guadagnini}. There, the CS potential is
integrated out, hence, $<W(C)>$ only depends on the curve $C$ .
Again, the metric independence of both the CS action and the
Wilson Loop leads to conclude that the result must be a knot (or
link) invariant.

Since we do not know how to solve the equation \eref{CS ecuacion}
exactly, we shall develop a perturbative solution. As it will be
seen, this procedure  leads to obtain the action on-shell as a
power series in $\Lambda$

\begin{equation}
S_{os}\,( [\gamma_i],\Lambda)= \sum_{p=0}^{\infty}\Lambda^p\,\,
S^{(p)}[\gamma_i] .\label{Sonshell}
\end{equation}
where $S^{(p)}[\gamma_i]$, the $p-th$ coefficient in the
expansion, carries the dependence on the curves $\gamma_i$. Now,
if $S _{os}\,( [\gamma_i],\Lambda)$ is a link invariant, so must
be their derivatives with respect to $\Lambda$. Hence, the
coefficients $S^{(p)}[\gamma_i]$ should be link invariants too. A
useful consequence of this simple argument, which is also valid
for the perturbative series of $<W(C)>$ in the quantum
case\cite{guadagnini}, is that one does not need to get the whole
power series in order to obtain link invariants. In this paper we
shall study the first two invariants that this method provide.

From equations  \eref{def I} and \eref{solucion I} we have

\begin{equation}
I_{i}({\tau})= U_{i}({\tau})g_{i}(0)K_{i}
g_{i}^{-1}(0)U_{i}^{-1}({\tau}) .\label{def I1}
\end{equation}
Hence, we can take $g_{i}({\tau})= U_{i}({\tau})g_{i}(0)$, which
implies
\begin{equation}
D_{\tau}g_{i}({\tau})= 0 ,
\end{equation}
and then
\begin{equation}
S_{os}^{interaction}=0.
\end{equation}
Thus, it remains to consider $S_{os}^{CS}$. To proceed further,
we find it  convenient to rewrite the equation of motion
\eref{conservacion} for the non-Abelian charge as

\begin{equation}
\frac{dI_{i}^{a}(\tau)}{d\tau}+ \Lambda
R_{i}^{ac}(\tau)I_{i}^{c}(\tau) = 0 ,\label{ecuacion IR}
\end{equation}
where we have defined

\begin{eqnarray}
R_{i}^{ac}(\tau)& \equiv&
R_{i\mu}^{ac}(z_{i}(\tau))\dot{z}_{i}^{\mu}(\tau) \nonumber\\
& \equiv & f^{abc} B_{i\mu}^{b}(\tau)\,\dot{z}_{i}^{\mu}(\tau),
\label{ecuaR}
\end{eqnarray}
with
\begin{equation}
 B_{\mu}= \Lambda^{-1}A_{\mu}.
\label{ecuaB}
\end{equation}
Solving equation \eref{ecuacion IR} we get

\begin{equation}
I_{i}^{a}(\tau)= \Big\{ \,T\,exp \,
\Big(-\Lambda^{-1}\,\int_{0}^{\tau} d\tau' R_{i}(\tau')
\Big)\Big\}^{ab}I_{i}^{b}(0)   ,\label{soluI}
\end{equation}
which is another form of writing the result given by eqs.
\eref{solucion I} and \eref{tiempo ord}. Introducing \eref{soluI}
in the equation of motion \eref{CS ecuacion}, and expanding the
time ordered exponential we arrive to the expression

\begin{eqnarray}
2\epsilon^{\mu\nu\rho}\,\partial_{\nu}B_{\rho}^{a}(x) & = &
-\Lambda
\,\epsilon^{\mu\nu\rho}\,f^{abc}B_{\nu}^{b}(x)B_{\rho}^{c}(x)
+\,\sum_{i=1}^{n} \,\oint_{\gamma_i} dz^{\mu}
\delta^{3}(x-z)\,I^{a}_{i}(0)\nonumber\\
& & -\Lambda \,\sum_{i=1}^{n}\,\oint_{\gamma_i}
dz^{\mu}\int_{0}^{z}
dz_{1}^{\mu_1}\,R^{aa_1}_{\mu_1}(z_1)\,\delta^{3}(x-z)\,I^{a_1}_{i}(0)\nonumber\\
& & +\Lambda ^{2} \,\sum_{i=1}^{n}\,\oint_{\gamma_i}
dz^{\mu}\int_{0}^{z} dz_{1}^{\mu_1}\int_{0}^{z_1}
dz_{2}^{\mu_2}R^{aa_1}_{\mu_1}(z_1)R^{a_{1}a_{2}}_{\mu_2}(z_2)\delta^{3}(x-z)\,I^{a_2}_{i}(0)\nonumber\\
& & \vdots \nonumber\\
& & +(-\Lambda )^{p}\sum_{i=1}^{n}\,\oint_{\gamma_i}
dz^{\mu}\int_{0}^{z} dz_{1}^{\mu_1}
\ldots\int_{0}^{z_{p-1}}dz_{p}^{\mu_p}R^{aa_1}_{\mu_1}(z_1)\ldots
R^{a_{p-1}a_{p}}_{\mu_p}(z_p) \delta^{3}(x-z)I^{a_p}_{i}(0)\nonumber\\
& & \vdots \label{maestra}
\end{eqnarray}
In this equation, we substitute $B_{\rho}^{a}$ by the power series

\begin{equation}
B_{\rho}^{a}= \sum_{p=0}^{\infty} \Lambda ^{p}
{B^{(p)}}_{\rho}^{a} ,\label{potenciasB}
\end{equation}
which allows us to write the equation that the $p-th$
contribution ${B^{(p)}}_{\rho}^{a}$ to the  potential obeys, in
the form

\begin{eqnarray}
2\epsilon^{\mu\nu\rho}\,\partial_{\nu}{B^{(p)}}_{\rho}^{a}(x)& =
& -\epsilon^{\mu\nu\rho}\,f^{abc}\sum_{r,s=0}^{r+s=p-1}
{B^{(r)}}_{\nu}^{b}(x){B^{(s)}}_{\rho}^{c}(x) +
\sum_{r=1}^{p}(-1)^{r}\sum_{i=1}^{n}\,\oint_{\gamma_i}dz^{\mu}\int_{0}^{z}dz_{1}^{\mu_1}\ldots\nonumber\\
\ldots \int_{0}^{z_{r-1}}& & dz_{r}^{\mu_r}  \sum_{s_{1},
\ldots,s_{r}=o}^{s_{1}+ \ldots + s_{r}=p-r}
{R^{(s_1)}}^{aa_1}_{\mu_1}(z_1){R^{(s_2)}}^{a_{1}a_{2}}_{\mu_2}(z_2)
\ldots {R^{(s_r)}}^{a_{r-1}a_{r}}_{\mu_r}(z_r)
\delta^{3}(x-z)\,I^{a_r}_{i}(0). \nonumber\\
\label{ecuap}
\end{eqnarray}
This equation holds for $p>0$. For $p=0$ one has, instead

\begin{eqnarray}
2\epsilon^{\mu\nu\rho}\,\partial_{\nu}{B^{(0)}}_{\rho}^{a}(x)& = &
\sum_{i=1}^{n}\,\oint_{\gamma_i}dz^{\mu}
\delta^{3}(x-z)\,I^{a}_{i}(0). \label{ecua0}
\end{eqnarray}
Despite its complicate appearance, equation \eref{ecuap} has two
nice features. Firstly, its right hand side involves $B^{(q)}$,
with $q<p$, hence, one can look for a recursive solution.
Secondly, the structure of eq.\eref{ecuap} is the same as that of
the $0-th$ order equation \eref{ecua0}: it is just like the
Ampere`s Law, whose solution is given by the Biot-Savart Law

\begin{eqnarray}
{B^{(p)}}_{\alpha}^{a}(x)= - \frac 1{4\pi} \int
dy^{3}\epsilon_{\alpha\beta\gamma} J^{(p)\beta
a}(y)\frac{(x-y)^{\gamma}}{|x-y|^{3}}\,+\,\partial_\alpha
f^{a}(x),\label{biot}
\end{eqnarray}
where $J^{(p)\beta a}(y)$ represents the r.h.s. of eq.
\eref{ecuap} (or \eref{ecua0}) divided by 2, with $f^{a}(x)$
being an arbitrary function that takes into account the
undeterminacy of the longitudinal part of
${B^{(p)}}_{\alpha}^{a}(x)$.

Once the equations of motion are solved perturbativelly, one must
consider the action on-shell, which may be written down as a power
series too

\begin{eqnarray}
S_{os} & = & S_{os}^{CS} \nonumber\\
& = & \frac\Lambda2 \int  d^3x \, \epsilon^{\mu\nu\rho} \,
(B_{\mu}^{a}
\partial_\nu B_{\rho}^{a}
+ \frac{\Lambda}{3} \, f^{abc}
B_{\mu}^{a} B_{\nu}^{b}B_{\rho}^{c})|_{on-shell}\nonumber\\
& = & \frac\Lambda2 \sum_{p=0}^{\infty} S^{(p)}\Lambda^{p}
\label{accion os1} ,
\end{eqnarray}
with

\begin{eqnarray}
 S^{(p)} & = & \int d^{3}x\,\epsilon^{\mu\nu\rho}\Bigg(\sum_{r,s}^{r+s=p}({B^{(r)}}_{\mu}^{a}
\partial_\nu {B^{(s)}}_{\mu}^{a}) \,+\,\frac13 f^{abc}\sum_{r,s,q}^{r+s+q=p-1}({B^{(r)}}_{\mu}^{a}
{B^{(s)}}_{\nu}^{b}{B^{(q)}}_{\rho}^{c})\Bigg), \label{Sp}
\end{eqnarray}
as can be verified after some algebra. From equations
\eref{ecuap}-\eref{Sp} we can obtain with a moderate effort the
first two contributions to  $S_{os}$. Firstly, we use
eq.\eref{biot} to write the solution of eq.\eref{ecua0} as

\begin{equation}
{B^{(0)}}_{\alpha}^{a}(x) = \sum_{i=1}^{n} D_{i\alpha}(x)
I_{i}^{a}(0), \label{B0}
\end{equation}
where we have defined
\begin{equation}
D_{i\alpha}(x) \equiv \frac1{4\pi} \oint_{\gamma_i} dz^{\gamma}
\frac{(x-z)^{\beta}}{|x-z|^{3}}\,\epsilon_{\alpha\beta\gamma}
\label{Dalex}.
\end{equation}
In the expression for ${B^{(0)}}_{\alpha}^{a}$ we have omitted the
gradient $\partial_\alpha f^{a}$, which does not contribute to
the  first two terms of $S_{os}$, as we shall see later. The
$0-th$ order contribution to $S_{os}$ is then given by
\begin{eqnarray}
S^{(0)}&=& \int d^{3}x
\epsilon^{\mu\nu\rho}{B^{(0)}}_{\mu}^{a}\partial_\nu
{B^{(0)}}_{\rho}^{a} \nonumber\\
& = & \frac14 \sum_{i,j}I_{i}^{a}(0)I_{j}^{a}(0)
L(i,j),\label{Accion0}
\end{eqnarray}
where
\begin{equation}
L(i,j) \equiv \frac1{4\pi}\oint_{\gamma_i}
dz^{\mu}\oint_{\gamma_j}
dy^{\rho}\frac{(z-y)^{\beta}}{|z-y|^{3}}\,\epsilon_{\mu\nu\rho},
 \label{Gauus}
\end{equation}
is the Gauss Linking Number (GLN) of $\gamma_i$ and $\gamma_j$. It
should be said that this expression  is not well defined when
$i=j$, although it can be converted into a meaningful expression
by applying certain regularization procedure, even in this
case\cite{guadagnini}. From eq.\eref{Accion0} it is evident that
the gradient $\partial_\alpha f^{a}$ does not contribute up to
this order. Also, we see that the first line in eq.\eref{Accion0}
is just the Abelian CS action, and the second one is precisely
the action OS obtained for the Abelian CS action coupled to
external particles that carry Abelian charges\cite{lorenzo2}.

Regarding the first order contribution to the action OS, one
finds, from the general results discussed above, the following
expression

\begin{equation}
S^{(1)}  =  \int
d^{3}x\,\epsilon^{\mu\nu\rho}\Bigg(2{B^{(0)}}_{\mu}^{a}
\partial_\nu {B^{(1)}}_{\rho}^{a} \,+\,\frac13 f^{abc}({B^{(0)}}_{\mu}^{a}
{B^{(0)}}_{\nu}^{b}{B^{(0)}}_{\rho}^{c})\Bigg). \label{accion1}
\end{equation}
Observe that ${B^{(1)}}_{\rho}^{a}$ enters in this expression (see
the first term) just through its rotational, which is given by eq.
\eref{ecuap} as

\begin{eqnarray}
\epsilon^{\mu\nu\rho}\,\partial_{\nu}{B^{(1)}}_{\rho}^{a}(x) &=&
-\frac 12 \,\epsilon^{\mu\nu\rho}\,f^{abc}
{B^{(0)}}_{\nu}^{b}(x){B^{(0)}}_{\rho}^{c}(x)\,- \nonumber\\
&-& \,\frac 12
\sum_{i=1}^{n}\,\oint_{\gamma_i}dz^{\mu}\int_{0}^{z}dz_{1}^{\mu_1}
{R^{(0)}}^{aa_1}_{\mu_1}(z_1)\delta^{3}(x-z)\,I^{a_1}_{i}(0).
\label{ecuap1}
\end{eqnarray}
Then, up to this order, we do not have to solve the equation for
${B^{(1)}}_{\rho}^{a}$ . Putting all together, we finally find

\begin{eqnarray}
S^{(1)}  =-\frac 14
\sum_{i,j,k}f^{abc}I^{a}_{i}(0)I^{b}_{j}(0)I^{c}_{k}(0)\,\Big\{\frac
13 \int d^{3}x
\,\epsilon^{\mu\nu\rho}D_{i\mu}(x)D_{j\nu}(x)D_{k\rho}(x)\,+
\nonumber\\
+\,\oint_{\gamma_i}dz^{\mu}\int_{0}^{z}dy^{\nu}D_{j\mu}(z)D_{k\nu}(y)\Big\}.\label{sigmaep}
\end{eqnarray}
This  expression vanishes when the isovectors $I^{a}_{i}(0)$,
$I^{b}_{j}(0)$, and $I^{c}_{k}(0)$ are linearly dependent. To
interpret our results, we shall consider the simplest
(non-trivial) case: let us assume that there are just three
particles of Wong, carrying independent isovectors at $\tau=0$.
Furthermore, let us also take $SU(2)$ as gauge group, and set
$I^{a}_{i}(0)= \delta_{i}^{a}$. Under these assumptions, $S^{(1)}$
can be writen as

\begin{eqnarray}
S^{(1)}(1,2,3)  &=& - \frac 12 \int d^{3}x
\,\epsilon^{\mu\nu\rho}D_{1\mu}(x)D_{2\nu}(x)D_{3\rho}(x)\, -\nonumber\\
&& - \frac 12 \int d^{3}x\int
d^{3}y\Big\{T_{1}^{[\mu x,\,\nu y]}D_{2\mu}(x)D_{3\nu}(y) + \nonumber\\
&& \qquad \qquad \qquad \quad  + T_{2}^{[\mu x,\,\nu y]}  D_{3\mu}(z)D_{1\nu}(y) + \nonumber\\
&& \qquad \qquad \qquad \quad + T_{3}^{[\mu x,\,\nu
y]}D_{1\mu}(z)D_{2\nu}(y) \Big\},\label{(1,2,3)}
\end{eqnarray}
where we have introduced the bilocal tensor density associated
with the curve $\gamma_i$

\begin{equation}
T_{\gamma_i}^{\mu x,\,\nu y} \equiv
\oint_{\gamma_i}dz^{\mu}\int_{0}^{z}dz'^{\nu}\delta^{3}(x-z)\delta^{3}(y-z').\label{tmunu}
\end{equation}
Observe that it is the antisymmetric part (in $\mu x,\,\nu y$) of
this object which enters in the expression for $S^{(1)}(1,2,3)$.
Also, $S^{(1)}(1,2,3)$ is antisymmetric under interchanges of the
curves 1,2,3.

Expression \eref{(1,2,3)} is, up to a factor, The Third Milnor
Linking Coefficient (TMLC)\cite{milnor}. We recognized it from
reference\cite{rozansky}, where this link invariant appears as a
contribution to the vacuum expectation value of the product of
Wilson Loops, in the context of perturbative Quantum non-Abelian
CS theory. The TMLC is a highly non-trivial link invariant
associated with three non-intersecting closed curves. It is
defined whenever the three curves do not link each other in the
Gauss sense, i.e.

\begin{equation}
L(i,j) \ne 0, \qquad \forall i,j,\qquad i\ne j. \label{Gaus0}
\end{equation}
In fact, the TMLC follows the Gauss Linking Number (which is then
the Second MLC) in an infinite sequence of link invariants
discovered by Milnor, the so called Higher Order Linking
Coefficients $K_n$. The $n-th$ coefficient makes sense only if
$K_p=0$, for $p<n$\cite{milnor}.

It is interesting to see how condition \eref{Gaus0} arises in our
scheme. First, observe that the $0-th$ order equation of motion
eq.\eref{ecua0} is trivially integrable, since its r.h.s. has
vanishing divergence. For the next order, (see eq.\eref{ecuap1})
the corresponding integrability condition (which again is
obtained by taking the divergence on both sides) is found to be

\begin{equation}
\sum_{i,j}f^{abc
}I^{b}_{i}(0)I^{c}_{j}(0)\,\delta^{3}(x-z_{i}(0))\int_{\gamma_i}dz^{\mu}
D_{j\mu}(z) = 0. \label{condicion1}
\end{equation}
Under the simplifications that lead to eq. \eref{(1,2,3)} (three
loops, $N=2$, and $I^{a}_{i}(0)= \delta_{i}^{a}$), eq.
\eref{condicion1} may be written as
\begin{equation}
\Big( \delta^{3}(x-z_{i}(0))- \delta^{3}(x-z_{j}(0))
\Big)L(i,j)=0, \label{condicion2}
\end{equation}
for any pair $i,j$, with $i\ne j$. If the curves do not intersect
each other, this equation just tells us that $L(i,j)=0$, as
expected. Thus, we obtain that the consistence condition under
which the first order action OS is meaningful, is precisely the
existence condition for the TMLC.

\section{General Covariance of the TMLC and consistence conditions}\label{sec4}
 Expression \eref{(1,2,3)} is not manifestly invariant under diffeomorphisms,
 since the kernel $\frac{(x-y)^{\mu}}{|x-y|^{3}}$ that enters in the definition of $D_{i\mu}(x)$ does not
 transforms covariantly under general coordinate changes.
 The same observation applies to expression \eref{Gaus0} for the GLN. It could be said that to solve the
 metric-independent equations of motion one has introduced a
 particular metric, the Euclidean one, that breaks the
 general covariance of the action OS. This can be better
 understood by analogy with gauge theories, where it is frequent to deal
 with non manifestly gauge-invariant expressions for gauge-invariant quantities, once the gauge has been
 fixed. In view of this it will be interesting to have generally
 covariant expressions for our link invariants, that allow us both to
 see explicitly that they are metric independent, and to dispose of an
 appealing interpretation of their geometrical meaning.

 With this goal in mind, we find it useful to define the
 following sequence of differential forms. To a volume $V$ of
 $R^3$, we can associate the $0-$ form

\begin{equation}
f(y,V) \equiv \int_{V}d^3x \,\delta^{3}(x-y), \label{0forma}
\end{equation}
 with support in $V$. Under a general coordinate transformation,
 the Jacobian that rises from the volume element compensates the inverse of the
 Jacobian produced by the Dirac's delta function. Hence, $f(y,V)$
 transforms covariantly under diffeomorphisms. As we shall see,
 this is a common feature of all the forms we are going to build up
 from $f(y,V)$. Also, it is worth observing that these forms are
 metric-independent. Taking the opposite of the exterior
 derivative of $f(y,V)$ we define the $1-$ form

\begin{eqnarray}
 g_{\mu}(y,\partial V) &\equiv& -\frac{\partial}{\partial y^\mu} f(y,V) \nonumber\\
 & = & \frac12 \int_{\partial V} d\Sigma^{\nu\rho}(x) \epsilon_{\mu\nu\rho}\,\delta^{3}(x-y), \label{1formadelta}
\end{eqnarray}
where  $\partial V$ is the boundary of $V$ and we have used
Stokes Theorem to produce the second line in the r.h.s. of this
equation. The $1-$ form $g(y,\partial V) =-df(y, V) $ is also
metric-independent and generally-covariant. Expression
\eref{1formadelta} also serves to define a $1-$ form
$g_{\mu}(y,\Sigma)$ for arbitrary (i.e., not necessarily closed)
surfaces $\Sigma$
\begin{eqnarray}
 g_{\mu}(y,\Sigma) &\equiv&
  \frac12 \int_{\Sigma} d\Sigma^{\nu\rho}(x) \epsilon_{\mu\nu\rho}\,\delta^{3}(x-y), \label{1formadeltageneral}
\end{eqnarray}
that enjoys the same transformation properties of
$g_{\mu}(y,\partial V)$. Taking the exterior derivative of this
object we obtain in turn

\begin{eqnarray}
 h_{\mu\nu}(y,\partial \Sigma) &\equiv& 2 \partial_{[\mu }g_{\nu ]}(y, \Sigma) \nonumber\\
 & = & \epsilon_{\mu\nu\rho} \oint_{\partial \Sigma}\, dx^{\rho}\delta^{3}(x-y),\label{2formadelta}
\end{eqnarray}
where we have employed Stokes Theorem again. The $2-$ form
$h(y,\partial \Sigma)=dg(y,\Sigma)$ can  also be extended to open
curves $\gamma$. In that case,
\begin{eqnarray}
 h_{\mu\nu}(y,\gamma) &\equiv&
 \epsilon_{\mu\nu\rho} \oint_{\gamma}\, dx^{\rho}\delta^{3}(x-y).\label{2formadeltageneral}
\end{eqnarray}
 From $h$ we can define the vector
density

\begin{equation}
T^{\mu y}_{\gamma}\equiv  \frac12  \epsilon^{\mu\nu\rho}\
h_{\nu\rho}(y,\gamma)= \oint_{\gamma}
dx^{\mu}\delta^{3}(x-y),\label{Tmu}
\end{equation}
that precedes to the bilocal density $T^{\mu x,\,\nu y
}_{\gamma}$ defined in eq. \eref{tmunu} in an infinite list of
"loop coordinates" with well studied properties\cite{gaetano}.
For our purposes, it suffices to notice that both objects are
metric-independent densities, and that $T^{\mu x,\,\nu y
}_{\gamma}$ obeys the "differential constraint"
\begin{eqnarray}
\frac{\partial}{\partial x^\mu}T^{\mu x,\,\nu y }_{\gamma} &=&
\Big(-\delta^{3}(x-x_{0})
+\delta^{3}(x-y)\Big)T^{\nu y}_{\gamma}\nonumber\\
\frac{\partial}{\partial y^\nu}T^{\mu x,\,\nu y }_{\gamma} &=&
\Big(\delta^{3}(y-x_{0}) -\delta^{3}(y-x)\Big)T^{\mu x
}_{\gamma},\label{diferencial}
\end{eqnarray}
and the "algebraic constraint"
\begin{equation}
T^{(\mu x,\,\nu y )}_{\gamma} \equiv \frac 12 \Big(T^{\mu x,\,\nu
y }_{\gamma} + T^{\nu y,\,\mu x }_{\gamma} \Big)= T^{\mu x
}_{\gamma} \, T^{\nu y}_{\gamma}.\label{algebraica}
\end{equation}

 To close the sequence, we take the opposite of the
exterior derivative of the $2-$ form  $h_{\mu\nu}(y, \gamma)$,
which defines then a $3-$ form with support on the boundary
$\partial\gamma$ as

\begin{eqnarray}
 i_{\mu\nu\rho}(y,\partial \gamma) &\equiv& -3 \partial_{[\mu }h_{\nu\rho ]}(y,\partial \gamma) \nonumber\\
 & = & \epsilon_{\mu\nu\rho} \Big(\delta^{3}(y-x_{f})-\delta^{3}(y-x_{0})\Big),\label{3formadelta}
\end{eqnarray}
with $x_{f}$ and $x_{0}$ being the starting and ending points of
$\gamma$. Finally, observe that, as in the previous cases, the
$3-$ form $i(y,\partial \gamma)= -dg(y,\gamma)$ may also be
trivially extended to the case of "open $0-$ spheres", i.e.,
single points

\begin{eqnarray}
 i_{\mu\nu\rho}(y,x) &\equiv&
  \epsilon_{\mu\nu\rho} \delta^{3}(y-x).\label{3formadeltageneral}
\end{eqnarray}
Now, let us consider the quantity

\begin{eqnarray}
I(\Sigma_{1},\Sigma_{2},\Sigma_{3})  &=& \int d^{3}x
\,\epsilon^{\mu\nu\rho}g_{1\mu}(x)g_{2\nu}(x)g_{3\rho}(x)\, +\nonumber\\
&& + \int d^{3}x\int
d^{3}y\Big\{T_{1}^{[\mu x,\,\nu y]}g_{2\mu}(x)g_{3\nu}(y) + \nonumber\\
&& \qquad \qquad \qquad \quad  + T_{2}^{[\mu x,\,\nu y]}  g_{3\mu}(x)g_{1\nu}(y) + \nonumber\\
&& \qquad \qquad \qquad \quad + T_{3}^{[\mu x ,\,\nu y
]}g_{1\mu}(x)g_{2\nu}(y) \Big\},\label{I(1,2,3)}
\end{eqnarray}
with $\Sigma_{1}$, $\Sigma_{2}$ and $\Sigma_{3}$ being arbitrary
surfaces. $g_{i\mu}(x)$ is just a shorthand for
$g_{\mu}(x,\Sigma_{i})$. Also, by $T_{i}^{[\mu x,\,\nu y]}$ we
mean the (antisymmetrized) bilocal density defined before,
evaluated at  the boundary $\partial\Sigma_{i}$ of $\Sigma_{i}$.
Due to the metric independence and general covariance of the
ingredients of expression \eref{I(1,2,3)}, it is immediate to see
that each one of its terms is a topological invariant of the
surfaces $\Sigma_{i}$. The first term measures how many times the
three surfaces intersect at a common point. The second term
counts the oriented number of times that the boundary
$\partial\Sigma_{1}$ crosses first the surface $\Sigma_{2}$ and
then $\Sigma_{3}$. The remaining terms have a similar
interpretation. Clearly, every one of these quantities is
invariant under continuous deformations of $R^3$. What it is by
no means trivial is to see whether or not they are link rather
than surfaces invariants. To study this point, let us compute how
$I(\Sigma_{1},\Sigma_{2},\Sigma_{3})$ changes when $\Sigma_{1}$
is replaced by another surface $\Sigma_{1}'$, such that both
surfaces share the same boundary: $\partial\Sigma_{1}$ =
$\partial\Sigma_{1}'$. One has
\begin{eqnarray}
\Delta I_{1} &\equiv& I(\Sigma_{1}',\Sigma_{2},\Sigma_{3}) -
I(\Sigma_{1},\Sigma_{2},\Sigma_{3})\nonumber\\
 &=& \int d^{3}x
\,\epsilon^{\mu\nu\rho} \Delta g_{1\mu}(x)g_{2\nu}(x)g_{3\rho}(x)\, +\nonumber\\
&& + \int d^{3}x\int d^{3}y \Big\{T_{2}^{[\mu x,\,\nu y]}  g_{3\mu}(x)\Delta g_{1\nu}(y) + \nonumber\\
&& \qquad \qquad \qquad \quad + T_{3}^{[\mu x,\,\nu y]}\Delta
g_{1\mu}(x)g_{2\nu}(y) \Big\}, \label{deltaI}
\end{eqnarray}
where
\begin{equation}
\Delta g_{1\mu} \equiv  g_{1\mu}(\Sigma_{1}')-
g_{1\mu}(\Sigma_{1})= g_{1\mu}(\partial V), \label{deltag1}
\end{equation}
$V$ being the volume enclosed by the surface that results of the
composition of $\Sigma_{1}'$ with the opposite of $\Sigma_{1}$:
$\partial V \equiv \Sigma_{1}' - \Sigma_{1}$. Now,
eq.\eref{1formadelta} tells us that $g(\partial V) = -df(V)$ .
Then we can write, after integrating by parts, the second term of
the second equality in eq. \eref{deltaI} as

\begin{eqnarray}
\int d^{3}x\int d^{3}y \frac{\partial}{\partial y^{\nu}
}\Big(T_{2}^{[\mu x,\,\nu y]}\Big) g_{3\mu}(x) f_{1}(y)&=&
f_{1}(x_{0}^{(2)}) \int d^{3}xT_{2}^{\mu x}g_{3\mu}(x) - \nonumber\\
 && \qquad -\int d^{3}xf_{1}(x)T_{2}^{\mu x}g_{3\mu}(x).\label{termino1}
\end{eqnarray}
In writing this equation, we have also employed the differential
constraints eq.\eref{diferencial} obeyed by $T^{[\mu x,\,\nu y
]}$. In a similar form, the last term of eq. \eref{deltaI} may
written as
\begin{eqnarray}
\int d^{3}x\int d^{3}y \frac{\partial}{\partial x^{\mu}
}\Big(T_{3}^{[\mu x,\,\nu y]}\Big) g_{2\nu}(y) f_{1}(x)&=&
-f_{1}(x_{0}^{(3)}) \int d^{3}xT_{3}^{\mu x}g_{2\mu}(x)  + \nonumber\\
 && \qquad +\int d^{3}xf_{1}(x)T_{3}^{\mu x}g_{2\mu}(x).\label{termino2}
\end{eqnarray}
Expressions \eref{termino1} and \eref{termino2} are related,
since in view of the definitions of $h_{\mu\nu}$ and $T^{\mu}$ one
has
\begin{eqnarray}
\int d^{3}xf_{1}(x)T_{3}^{\mu x}g_{2\mu}(x) & = &\int
d^{3}xf_{1}(x)
\epsilon^{\mu\nu\rho}\partial_{\nu}g_{3\rho}(x)g_{2\mu}(x)
\nonumber\\
&=& \int d^{3}x \Big[ -\epsilon^{\mu\nu\rho}\Delta
g_{1\mu}(x)g_{2\nu}(x)g_{3\rho}(x)+ \int d^{3}xf_{1}(x)T_{2}^{\mu
x }g_{3\mu}(x) \Big] \label{ter12}.
\end{eqnarray}
Then, substituting eqs. \eref{termino1} and \eref{termino2} into
\eref{deltaI}, and using the identity \eref{ter12} we finally get
\begin{equation}
\Delta_1 I(\Sigma_1,\Sigma_2,\Sigma_3) = \Big(f_{1}(x_{0}^{(2)}) -
f_{1}(x_{0}^{(3)})\Big) L(\partial\Sigma_2 , \partial\Sigma_3)
\label{deltaIfinal},
\end{equation}
where we have also employed that the GLN $L(\partial\Sigma_2 ,
\partial\Sigma_3)$ is equal to the "crossing number" of
$\partial\Sigma_2$ with $\Sigma_3$, and that it is symmetric under
exchange of $\Sigma_2$ with $\Sigma_3$\cite{rolfsen}
\begin{eqnarray}
L(\partial\Sigma_2 ,\partial\Sigma_3) =\int d^{3}x T_{2}^{\mu x}
g_{3\mu}(x) = \int d^{3}x T_{3}^{\mu x} g_{2\mu}(x)
\label{linkflux}.
\end{eqnarray}
Hence, from eq. \eref{deltaIfinal} we find that $ \Delta_1 I$
vanishes provided that the curves 2 and 3 do not intersect and
have vanishing GLN. These  arguments can be obviously repeated to
calculate the dependence of $I(\Sigma_{1},\Sigma_{2},\Sigma_{3})$
on the surfaces $\Sigma_{2}$ and  $\Sigma_{3}$, with similar
conclusions. The result is then that
$I(\Sigma_{1},\Sigma_{2},\Sigma_{3})$  is a link invariant of the
curves that bound the surfaces $\Sigma_{i}$ (besides being a
"surfaces-invariant" quantity) provided that these curves does not
cross each other and have vanishing GLN.

This result, together with the appearance of
$I(\Sigma_{1},\Sigma_{2},\Sigma_{3})$ leads us naturally to ask
which is the relation, if any, between this quantity and the
first order contribution to $S_{os}$ (eq.\eref{(1,2,3)}) obtained
in the previous section. To seek for the precise relation between
these quantities, let us observe that the key for establishing the
independence of $I(\Sigma_{1},\Sigma_{2},\Sigma_{3})$ of the
surfaces, was the fact that it does not vary when  the
$g_{\mu}(\Sigma)`s$ change by  additive gradients
$\partial_{\mu}\Lambda$ (provided the GLN's of the boundaries
$\partial\Sigma`s$ vanish). But a direct calculation shows that

\begin{equation}
\partial_{\mu}D_{i\nu}(x) - \partial_{\nu}D_{i\mu}(x) = \partial_{\mu}g_{i\nu}(x) - \partial_{\nu}g_{i\mu}(x)
\label{dequalg},
\end{equation}
thus, $D_{i\nu}$ and $g_{\mu}$ just differ by a gradient and one
may replace the latter by the former in expression
\eref{I(1,2,3)}. Hence, one has
\begin{equation}
S^{1}(\partial\Sigma_{1},\partial\Sigma_{2},\partial\Sigma_{3})=
-\frac 12 I(\Sigma_{1},\Sigma_{2},\Sigma_{3}) + \Delta
\label{SequalI},
\end{equation}
where $\Delta$ is a function of the surfaces that vanishes when
the GLN are equal to zero. This equation provides the relation we
were looking for. In passing, we see that by the same argument it
is allowed to neglect the gradient $\partial_{\alpha}f^{a}$ that
comes from the "Biot-Savart" solution  also to compute the first
order contribution to $S_{os}$.

The result we have obtained in this section could be summarized
as follows. It is possible to provide two equivalent analytical
expressions for the TMLC of a set of three curves. One of them is
not manifestly invariant under diffeomorphisms and is given
explicitly by $-2S^{1}(1,2,3)$. In turn, the other one, given by $
I(\Sigma_{1},\Sigma_{2},\Sigma_{3})$ is manifestly invariant under
diffeomorphisms, but it is not explicitly "link-dependent";
instead, it is "surface-dependent". Both expression are related
through a geometric mechanism: changing from the "link" to the
"surfaces" presentation, amounts to performing an "Abelian gauge
transformation" under which the TMLC is invariant.

\section{Discussion}

We have presented a method for obtaining link invariants through
the study of the classical equations of motion of non-Abelian
Chern-Simons theory coupled to linked sources. The method relies
on the fact that the classical action that we take should retain
its topological character when it is calculated on-shell.
Furthermore, a simple argument allows to see that this is true
even perturbativelly. We have studied the first two invariants
that the method provides. While the first one is rather trivial
(in the sense that it appears in almost all the discussions about
link invariants and Chern-Simons theory), the second one is
highly non-trivial, and corresponds to the Third Milnor Linking
Coefficient \cite{milnor}. This invariant is useful, for instance,
to characterize the entanglement properties of the Borromean Rings
\cite{milnor,rolfsen}, which constitute a non-trivial
three-component link that has vanishing Gauss Linking Number
between any pair of its components.

We have also introduced a geometrical setting that allows us to
write down the TMLN in a manifestly diffeomorphism invariant
form, although in this presentation this object looks like a
"surfaces" rather than a "link"  invariant. The surfaces
appearing in this presentation are such that their boundaries are
the components of the link (they are Seifert Surfaces, in
knot-theoretical parlance). This fact, instead of being an
inconvenience, is welcomed: it allows to interpret the TMLC as a
particular combination of intersection numbers between the
surfaces and their boundaries. It is interesting to point out
that there is a recent work \cite{interpretacion} devoted to the
interpretation of the TMLC, with which our results should be
compared.

The explicit choice of  a particular set of Seifert Surfaces in
the expression for the TMLC is seen to be related with a kind of
Abelian Gauge Symmetry: the surfaces enter in that expression
through certain $1-$forms. Changing a Seifert Surface by another
one, amounts to shifting its associated $1-$form by a gradient,
and this transformation is seen to leave the "surface-dependent"
expression unchanged; thus, one obtains that the dependence in
the surfaces is realized only through their boundaries, as
corresponds to a link-invariant.

This work was supported by \emph {Consejo de Desarrollo
Cient\'{\i}fico y Human\'{\i}stico}, Universidad Central de
Venezuela, Caracas, VENEZUELA.

\end{document}